\documentclass[twocolumn,showpacs,preprintnumbers,amsmath,amssymb]{revtex4}

\usepackage{graphicx}
\usepackage{dcolumn}
\usepackage{bm}

\begin{document}

\preprint{submitted to Supercond. Sci. Technol.}

\title{Direct visualization of iron sheath shielding effect in MgB$_2$ superconducting wires}

\author{Alexey V. Pan}
\email{pan@uow.edu.au}
\author{Shihai Zhou}
\author{Huakun Liu}
\author{Shixue Dou}
\affiliation{Institute for Superconducting and Electronic Materials, University of Wollongong, \\ Northfields Avenue, Wollongong, NSW 2522, Australia}

\date{3 May 2002, Revised 9 May 2003}

\begin{abstract}
Local magneto-optical imaging and global magnetization measurement techniques were used in order to visualize shielding effects in the superconducting core of MgB$_2$ wires sheathed by ferromagnetic iron (Fe). The magnetic shielding can provide a Meissner-like state in the superconducting core in applied magnetic fields up to $\sim 1$~T. The maximum shielding fields are shown to correlate with the saturation fields of magnetization in Fe-sheaths. The shielding has been found to facilitate the appearance of an overcritical state, which is capable of achieving a critical current density ($J_c$) in the core which is larger than $J_c$ in the same wire without the sheath by a factor of $\sim 2$. Other effects caused by the magnetic interaction between the sheath and the superconducting core are discussed.
\end{abstract}
 
\maketitle

 New promising horizons for applications of superconducting wires were opened up by the discovery of MgB$_2$ superconductor \cite{discovery}. MgB$_2$ wires have been reported to have high values of the critical current density $J_c$, in particular in wires sheathed by mechanically hard and magnetically soft materials such as iron (Fe) \cite{said}. In spite of an initial fear of local superconductivity suppression in the vicinity of the sheath-core interface, the ferromagnetic iron sheath has been found to cause a magnetic shielding effect which provided an {\it extended} range of a constant critical current $I_c$ in an externally applied magnetic field $B_a$, $0.2 \le B_a \le 0.6$~T at $T = 32$~K \cite{joseph}. The same screening effect led to either negligibly small or significantly reduced AC losses in the sheathed wires, in contrast to the high level of the losses in the wires without the screening \cite{sumpt2}. However, it is yet unclear to what extent this shielding affects the superconducting behaviour and critical currents in the wires. In fact, the high permeability of the sheath, enabling the screening \cite{classic}, prevents magnetic flux entry into the superconducting core. This means that the magnetic field range of the Meissner state can be effectively extended (or shifted) to higher $B_a$ (a pseudo-Meissner effect \cite{sumpt}). Moreover, the interaction between soft magnetic and superconductor can enable an overcritical state carrying $J > J_c$ \cite{gen} {\it without introducing additional pinning centers}. In this work, by employing magneto-optical (MO) imaging and magnetization measurement techniques we directly visualize the iron shielding in MgB$_2$ wires and show a number of effects that are the consequences of this shielding.

\begin{figure}[t]
\includegraphics[scale=0.41]{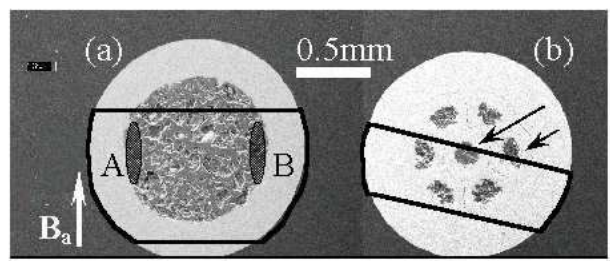}
\caption{\label{cross}Cross section of the mono-core (a) and multifilamentary (b) wires. The black lines indicate the remaining parts after polishing the longitudinal sections for MO imaging. The $B_a$-vector shows the direction of the applied field for the MC wire. A and B shadowed areas show the regions of the large Meissner shielding super-currents.}
\end{figure}

Fe-sheathed MgB$_2$ mono-core (MC) and multi-filamentary (MF) superconducting wires were investigated. A detailed description of the wire manufacturing procedure and its characterization is given elsewhere \cite{pan}. The MC and MF wire cross sections are shown in Fig.~\ref{cross}. Transport measurements of the wires provided the critical currents $I_c = 78.1$~A and 110~A at temperature $T = 4.2$~K in self-field for the MC and MF wires, respectively. To obtain a magnetic flux distribution over the wires with the help of the MO technique, the samples were polished down to the final longitudinal cross sections which are indicated by the black lines in Fig.~\ref{cross}. In the case of the MF wire, only two filaments denoted by arrows were exposed for taking MO images. An epitaxial ferrite-garnet magneto-optical indicator film with in-plane magnetization was used for the imaging. Since both kinds of the wires have exhibited a similar qualitative picture, MO observations will be shown only for the MC wire. Global magnetization measurements as a function of the applied field within $|B_a| \le 5$~T at different temperatures were performed with the help of a Quantum Design MPMS SQUID magnetometer. The field was applied perpendicular to the polished planes of the wires for both, local and global, kinds of measurements.

\begin{figure}
\vspace{0.5cm}
\includegraphics[scale=0.38]{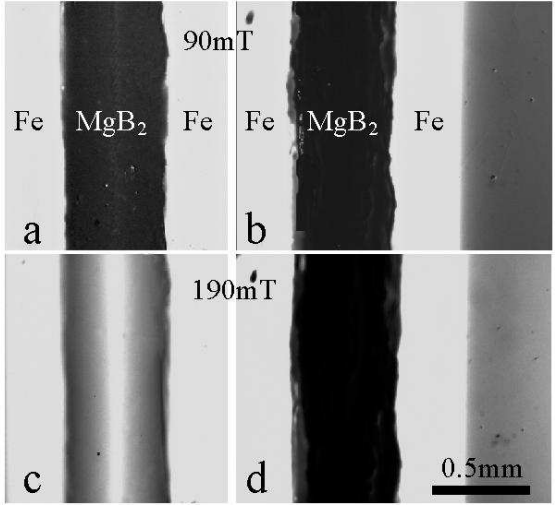}
\caption{\label{MC}Magneto-optical images of the Fe-sheathed MC wire at RT (a, c), and at $T = 19$~K (b, d) in the magnetic fields as denoted on each image row.}
\end{figure}

Two sets of MO images shown in Fig.~\ref{MC} were taken at different $B_a$ in the normal state at room temperature (RT) and in the zero-field cooled superconducting state at $T = 19$~K. The critical temperature $T_c$ of the samples is $\simeq 38.6$~K. At $B_a \le 90$~mT, there is almost no differences between the MO images taken in the normal (Fig.~\ref{MC}(a)) and superconducting (Fig.~\ref{MC}(b)) states, except in the vicinity of the sheath-core interface. Generally, the darker the area on the MO images, the lower the magnetic flux density $B_i$ in that region; the brighter the area, the higher the flux. As can be seen, the flux is the highest in the Fe-sheath ($B_i > B_a$) and the lowest within the core ($B_a > B_i \sim 0$).

At $B_a > 90$~mT, discrepancies between images taken at RT and 19~K (Fig.~\ref{MC}c, d) become obvious. The bright stripe of the flux in the middle of the core at RT and $B_a = 190$~mT is due to the geometry of the open sheath cavity (Fig.~\ref{cross}a) and because in the normal state, MgB$_2$ material is ``transparent" to the magnetic flux, having no influence on its distribution. On the contrary, the stripe is absent in the images taken in the superconducting state at $T = 19$~K, because the core is strongly diamagnetic, in particular, below the lower critical field $B_{c1} = (B_c \ln\kappa)/(\kappa\sqrt{2})$, where $\kappa$ is the Ginzburg-Landau parameter and $B_c$ - the thermodynamic critical field. In MgB$_2$ $B_{c1} \sim 60$~mT \cite{bc1}. Therefore, the magnetic flux is expelled from the interior of the core $B_i \simeq 0$~T, so that no stripe is observed (Fig.~\ref{MC}d). In spite the fact that the wire is placed in $B_a > B_{c1}$, the core remains entirely diamagnetic at $T < T_c$, with the exception of some regions along the sheath (Fig.~\ref{MC}b, d). In these regions $B_i \ne 0$ could be either due to some trapped flux in cracks and voids in the core or, most probably, due to Meissner shielding currents. The shielding currents in the vicinity of the sheath-core interface arise because of the geometry of the iron sheath cylinder which, in contrast to the complete screening case of a soft magnetic sphere or infinite cylinder \cite{classic}, was polished to open the core for the MO experiments. The fact that at $B_a > B_{c1}$ and $T < T_c$, $B_i$ within the core is $\sim 0$~T, we refer to as the extended Meissner-like state enabled by the shielding.

\begin{figure}
\includegraphics[scale=0.45]{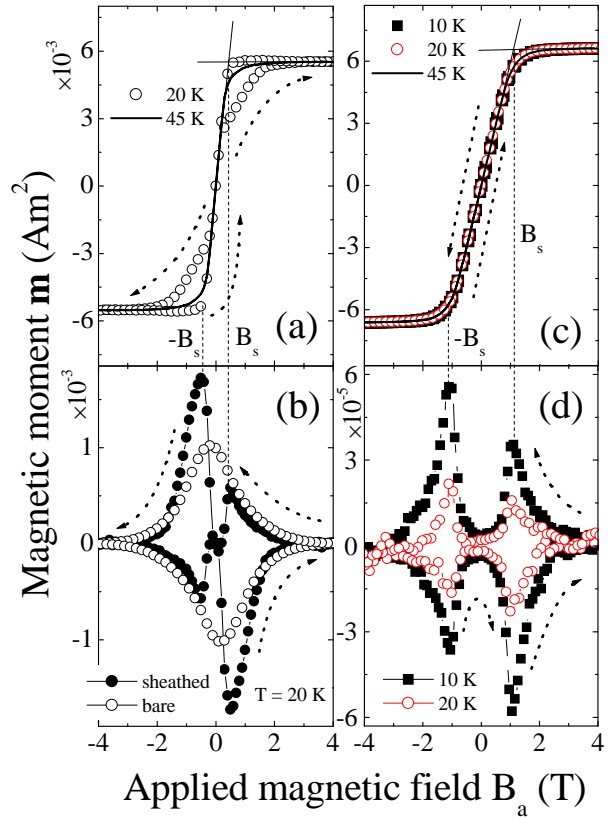}
\caption{\label{both}Magnetic moment curves of the MC wire measured at $T = 20$ and 45~K (a), and of the MF wire measured at $T = 10$, 20, 45~K (c). (b) Magnetic moment behaviour of the Fe-sheathed MC wire at $T = 20$~K after subtracting the Fe-sheath ferromagnetic contribution (solid symbols) and the magnetic moment measured after having the sheath removed (open symbols). (d) Magnetic moment of the MF wire at $T = 10$ and 20~K after subtracting the contribution of the sheath. Note the {\bf m}-scale differences.}
\end{figure}

Global magnetization measurement results are shown in Fig.~\ref{both} for both MC and MF wires. Figs.~\ref{both}(a) and (c) show curves measured for the {\it same} Fe-sheathed wires investigated by the MO technique. The ferromagnetic signal of the sheath dominates for both types of the wires below and above $T_c$. If we now subtract the sheath contribution measured above $T_c$ at 45~K from the curves measured below $T_c$, we obtain the superconducting contribution shown in Fig.~\ref{both}(b) by solid circles for $T = 20$~K for the MC wire and in (d) for 10~K and 20~K for the MF wire. A striking feature observed in these curves is the strongly suppressed (vanishing) signal at low applied fields: $|B_a| \le 0.5$~T for the MC wire and $|B_a| \le 1$~T for MF wire. These field intervals we attribute to the full operational range of the extended Meissner-like state in the wires. Apparently, the ranges go far beyond the maximum field available for our MO experiments ($B_a^{\rm max} \simeq 0.2$~T). Note that the region of the suppressed magnetic moment is independent of temperature (Fig.~\ref{both}(d)), indicating its non-superconducting origin. Instead, this region is well correlated with the ferromagnetic response of iron, showing the saturation of the magnetic moment at $|B_s| = 0.42$~T and 1.12~T for the MC and MF wires, respectively. $B_s$ is defined at the crossing of the lines as shown in Figs.~\ref{both}(a) and (c). The saturation occurs when the magnetic domains in the iron become aligned along the applied magnetic field. Below $|B_s|$ the flux is accumulated in the highly permeable Fe-sheath. Therefore, the filaments are effectively screened from the applied field. Above $|B_s|$ the screening is no longer complete. The smaller amount of the Fe-sheath material around the MC wire and the weaker stress in the sheath caused by the fabrication procedure are likely to be the factors responsible for the smaller $B_s$ value and, correspondingly, the narrower field range of the shielding compared to the MF wire. For the field applied along the axis of the wires we have measured \cite{pan2} one order of magnitude smaller $B_s$ values and narrower shielding ranges than in the transverse case presented in this work. This can be explained by the mechanical processing of the wires which creates an easy-axis in the Fe-sheath along the direction of the wire drawing.

To visualize the difference in the global magnetic behaviour of the wires which is provided by the Fe-shielding, we carefully removed the Fe-sheath from the MC wire so that the superconducting volume of the core remained {\it unchanged} after the removal. Magnetization measurements were carried out on this {\it bare} superconducting core. In Fig.~\ref{both}(b) we compare two magnetic moment curves for the Fe-sheathed and bare superconducting wires measured at $T = 20$~K. Strikingly, the magnetic moment of the Fe-sheathed wire can be larger than that for the bare wire by a factor of up to about 2 for {\it increasing} $|B_a|$. The magnetic moment of the sheathed wire becomes larger than that of the bare wire at $B_a \simeq B_s$ and remains larger up to $B_a \simeq 1.5$~T. Since the width of a superconducting magnetic moment loop is proportional to the critical current density $J_c$, it is straightforward to conclude that the Fe-sheath can provide a significantly higher $J_c$ than that in the wires without shielding for certain fields. This is an analog of the overcritical state discussed in Ref.~\cite{gen} for superconducting strips. In the case of the strips the magnetic interaction between certain geometries of a soft magnetic environment and a superconducting strip led to super-current redistribution over the strip resulting in a significant total critical current enhancement \cite{gen}. In our case, the soft magnetic environment, which is the Fe-cylinder (sheath), interacts with the superconducting MgB$_2$ core. The flux distribution over the strip with a transport current applied \cite{gen} and over the round wire with the field applied transversely to the wire \cite{sumpt} are similar, enabling the comparison. $B_i$ is higher at the sides than in the middle for both cases. This implies that at $B_a$ slightly larger than $B_s$, the Meissner shielding super-currents, flowing perpendicular to $B_a$, should be expected to be stronger at the sides of the core (A and B areas marked in Fig.~\ref{cross}(a)). As in the case considered for the strips \cite{gen}, the magnetic interaction between the sheath and the superconducting core redistributes the super-currents so that an overcritical state is achieved (Fig.~\ref{both}(b)). In contrast to the strips, magnetically sheathed wires may offer a simpler technological approach to the fabrication of long superconductors with {\it magnetically} enhanced current-carrying abilities. It is worth noting that in the case of the field applied parallel to the wire, no overcritical state was measured \cite{pan2}. In the parallel case, the shielding super-currents are evenly distributed within the core with the well-known Bean-like profile which is similar to that in the bare wire at the same conditions. Therefore, no redistribution should be expected.

Above 1.5~T in {\it increasing} $|B_a|$, the magnetic moment of the Fe-sheathed wire appears to be slightly suppressed compared to the bare wire (Fig.~\ref{both}(b)). This can be explained as follows. The sheath, being a permeable material, accumulates $B_i$ that is higher than $B_a$. Therefore, as the superconductivity weakens in the increased field, one can expect a rapid magnetization drop in fields approaching to the irreversibility field $B_{\rm irr}$, defined at the merger of the descending and ascending magnetic moment branches. Eventually (above $B_a \simeq 1.5$~T), the magnetized sheath starts to suppress the superconductivity inside the core. The magnetic moment of the sheathed wire is also slightly suppressed in {\it decreasing} $|B_a|$ up to $\sim B_s$. This is also caused by the magnetized sheath having $B_i > B_a$, which along with pinning delays the exit of the frozen flux from the core. This magnetic history dependence is in agreement with the prediction of \cite{gen}. As a result of the superconductivity suppression, $B_{\rm irr}$ is smaller in the sheathed sample than in the bare wire. Indeed, $B_{\rm irr}^{\rm sheathed} \simeq 3.0$~T, whereas $B_{\rm irr}^{\rm bare} \simeq 4.7$~T.

In conclusion, we visualized both locally and globally the magnetic shielding of the superconducting MgB$_2$ filaments in the MC and MF wires sheathed by iron. A Meissner-like state with $B_i \sim 0$~T can exist up to the saturation field of the ferromagnetic iron. Depending on the Fe-sheath properties, $B_s$ was measured as large as $\sim 1.12$~T. The overcritical state observed in the sheathed wire can have $J_c$ twice as large as in the same wire after removal of the sheath. However, $B_{\rm irr}^{\rm sheathed} < B_{\rm irr}^{\rm bare}$. Unlike the bare wire, the magnetization of the sheathed wire strongly depends on magnetic history.

\begin{acknowledgments}
We thank J. Horvat and T. Silver for careful reading of the manuscript and critical remarks. We also benefited from fulfilling discussions with Yu. A. Genenko. This work is financially supported by the Australian Research Council.
\end{acknowledgments}

\end{document}